\documentclass[11p]{article}
\usepackage{amsfonts,amsmath,amssymb,amscd,float}
\usepackage{graphicx}
\usepackage{authblk}
\usepackage[numbers,sort&compress]{natbib}
\providecommand{\keywords}[1]{\textit{\textit{Keywords---}} #1}
\title{Maxwell, Yang-Mills, Weyl and eikonal fields defined by any null shear-free congruence}

\author[1]{Vladimir.~V.~Kassandrov\thanks{vkassan@sci.pfu.edu.ru}}
\author[2]{Joseph.~A.~Rizcallah \thanks{joeriz68@gmail.com}}
\affil[1]{Institute of Gravitation and Cosmology, Peoples' Friendship University of Russia, Moscow, Russia}
\affil[2]{School of Education, Lebanese University, Beirut, Lebanon}

\begin{document}

\maketitle

\begin{abstract}
We show that (specifically scaled) equations of shear-free null geodesic congruences on the 
Minkowski space-time possess intrinsic self-dual, restricted gauge and algebraic structures. The  
complex eikonal, Weyl 2-spinor, $SL(2,\mathbb C)$ Yang-Mills and complex Maxwell fields, the latter produced by 
integer-valued electric charges (``elementary'' for the Kerr-like congruences), can all be explicitly associated with any shear-free null geodesic congruence. Using twistor variables, we derive the general solution of the equations of the shear-free null geodesic congruence (as a modification of the Kerr theorem) and analyze the corresponding ``particle-like" field distributions, with bounded singularities of the associated physical fields. These can be obtained in a straightforward algebraic way and exhibit non-trivial collective dynamics simulating physical interactions.
\end{abstract}


{\footnotesize PACS (2010): 02.40.-k, 03.50.-z, 02.10.D}

\section{Introduction and motivation}
\label{sec1}
It is well known that the structure of null geodesic congruences is
deeply related to the geometry and physics of space-time itself (see, e.g.
~\cite{pr,pr00,LeBrun}). In particular, {\it shear-free} null congruences
~\cite{pr,pr00,NewmanSFC,Trish} stand out for a number of remarkable properties, even on a
flat Minkowski background $\bf M$. For example, numerous field structures, electromagnetic in particular, can be associated with a shear-free congruence~\cite{pr,prmc,rob,NewmanSFMaxwell,kass9}.
Moreover, the geodesic and shear-free conditions are both invariant under
deformations of the metric of the Kerr-Shild type. If the congruence is
shear-free, then the entire system of vacuum Einstein or Einstein-Maxwell
equations can often be satisfied with an appropriate choice of a single 
``potential" function. In this way, an important Kerr-like class of (algebraically special)
solutions can be constructed~\cite{dks, demyan,burin}.

On the other hand, the equations of a shear-free null geodesic congruence possess a natural {\it twistor structure}. In
fact, these equations have the form~\footnote{Throughout the paper the standard two-component spinor
formalism is used. Upper case Latin indices range and sum over $0$ and $1$,
and are raised and lowered by the symplectic spinors  $\epsilon^{AB}$ and
$\epsilon_{AB}$ respectively, as in ~\cite{pr}. The symbol $\nabla_{AA^\prime}$ denotes the
usual spinor derivative operator in Minkowski space.}
\begin{equation}\label{shearfr}
\xi^{A^\prime} \xi^{B^\prime} \nabla_{A A^\prime} \xi_{B^\prime} = 0,
\end{equation}
where the 2-spinor $\xi(x)$ corresponds to the tangent 4-vector field of the
congruence $k_\mu=\xi_A\xi_{A^\prime}$. According to the celebrated {\it
Kerr theorem} ~\cite{pr,dks}, the general (analytical) solution of
(\ref{shearfr}) can be obtained in an implicit algebraic form
\begin{equation}\label{pi1}
\Pi (T) \equiv \Pi (\xi, iX\xi) = 0,
\end{equation}
where $\Pi$ is a {\it homogeneous} holomorphic function of twistor components
$T =\{\xi_{A^\prime}, \tau^A\}$ while the spinors $\xi, \tau$ are related
to the hermitian matrix of space-time coordinates $X = \{X^{A A^\prime}\}$ via
the {\it incidence condition}
\begin{equation}\label{tv}
\tau = iX\xi ~~~~(\tau^A = iX^{A A^\prime} \xi_{A^\prime}).
\end{equation}
Since both (\ref{shearfr}) and (\ref{pi1}) are invariant under generic local
rescalings of the spinors $\xi \mapsto \alpha(x) \xi,~~\tau \mapsto \alpha(x)\tau$, in
(\ref{pi1}) we are actually dealing with an {\it arbitrary} holomorphic function $\Pi$
of three {\it projective} twistor coordinates geometrically defining a smooth 2-surface in $\mathbb{C}P^3$~\cite{espos}.

However, there exists another, in essence equivalent to (\ref{pi1}), form of the Kerr theorem~\cite{kr1,kr4} which can be obtained via  
the transition to the {\it full twistor space} $T$ and explicitly 
exhibits a number of non-canonical symmetries, algebraic and
geometric structures, themselves leading to a self-induced field and particle-like dynamics. 
Those, rather unfamiliar properties of the shear-free null geodesic congruence, will be explored in this paper. 

Briefly, consider a smooth 2-surface in a 4-dimensional twistor
space $T$ defined by two algebraic equations
\begin{equation}\label{sol}
\Pi^{(C)} (T) \equiv \Pi^{(C)} (\xi, iX\xi) = 0, ~~~(C=0,1),
\end{equation}
with $\Pi^{(C)}$ being two {\it arbitrary} and independent holomorphic functions of the four twistor coordinates. For a couple of functions $\Pi^{(C)}$ the system
(\ref{sol}) determines locally on $\bf M$ a 2-spinor field $\xi(x)$.
Differentiating (\ref{sol}) w.r.t. the coordinates $X^{A
A^\prime}$, we get for the spinor (except at the singular points, see below)
\begin{equation}\label{rbccv}
\xi^{A^\prime} \nabla_{A A^\prime} \xi_{B^\prime} = 0,
\end{equation}
which, compared with (\ref{shearfr}), evidently defines a shear-free null geodesic congruence and, moreover, under special scaling of $\xi(x)$ is actually equivalent ~\cite{Trish} to its equations (\ref{shearfr}) .  
In turn, (\ref{rbccv}) defines a new spintensor $\Phi_{A^\prime A}$~\cite{Poten} since it can be equivalently represented in the form
\begin{equation}\label{delx}
\nabla_{A A^\prime} \xi_{B^\prime} = \Phi_{B^\prime A} \xi_{A^\prime} ~.
\end{equation}

Finally, we observe that the last system (\ref{delx}) admits an invariant Pfaffian (matrix) form
~\cite{kass5,kass1,kass2}
\begin{equation}\label{GSE}
\boxed{d\xi=\Phi dX \xi}
\end{equation}
for a 2-spinor $\xi(x)=\{\xi_{A^\prime}\}$ and a $Mat(2,\mathbb{C})$-valued
vector field $\Phi(x)=\{\Phi_{A^\prime A}\}$. In the case of flat space-time, we have
$dX=\{dX^{A A^\prime}\}$ and the symbol $d$ in the left-hand side of (\ref{GSE})
denotes the ordinary operator of external differentiation. System (\ref{GSE}), its solutions and related geometric and physical structures are the main concern of the present paper.  

Geometrically, system (\ref{GSE}) defines a covariantly-constant spinor field with respect to an effective matrix-valued connection 1-form
$\Omega=\Phi dX$. Spinor connections of similar type (on a generic Riemannian
background) have been considered by K.P. Tod~\cite{Tod} in their relations to self-dual Einstein-Weyl
manifolds and other geometric structures.

System (\ref{GSE}) is overdetermined, so both the 2-spinor
and the gauge fields are to be found from it in a self-consistent manner.
The integrability conditions of (\ref{GSE}) take the form
\begin{equation}\label{comp22}
dd\xi \equiv 0 = (d\Phi - \Phi dX \Phi)\wedge dX \xi \equiv R\xi
\end{equation}
and, as will be shown below, lead to self-duality of the curvature 2-form $R$
of the connection $\Omega$. Consequently, {\it the source-free Maxwell and Yang-Mills
equations are satisfied identically} on the solutions to (\ref{GSE}), for the
trace and trace-free parts of the curvature $R$ respectively. For this
reason, in what follows system (\ref{GSE}), equivalent to (\ref{delx}), (\ref{rbccv}) and, actually, to the defining equations of shear-free null geodesic congruence (\ref{shearfr}),  is
referred to as the {\it generating system of equations}. Thus, the source-free Maxwell, the complex Yang-Mills and the Kerr-Schild
metric fields can all be naturally associated with the solutions of the generating system of equations (\ref{GSE}). 

Moreover, the singularities of the strengths (curvatures) of
all these fields coincide in space and time, being all determined by a single
condition~\footnote{The Yang-Mills field possesses additional string-like singularities, see Section 5}
\begin{equation}\label{sinold}
\det \Vert \frac{d \Pi^{(C)}}{d\xi_{A^\prime}} \Vert = 0,
\end{equation}
which specifies the points where equations (\ref{sol}) admit multiple solutions. Geometrically, these points constitute a {\it caustic} of the corresponding shear-free light bundle.
As we shall see, this singular locus may be point-like,
string-like or even a 2-dimensional surface. In the case when the set is bounded
in 3-space we can identify it with a {\it particle-like object} whose time evolution
is fully  governed by the field equations (\ref{GSE}) and may be obtained
from the algebraic system (\ref{sol}) along with the condition (\ref{sinold}).

These singular objects exhibit, at a purely classical level,
some properties of the real quantum particles. In particular, the value of
electric charge $q$ is either zero or a whole multiple of the charge of the
fundamental static solution; the latter corresponds just to the Kerr shear-free
congruence with a ring-like singularity. The associated metric and electromagnetic
field are identical to those of the Kerr-Newman solution in the general theory of relativity,
except for the restrictions on the admissible value of the electric charge~\footnote{In the dimensionless units we use; the numerical value
itself is of no significance.} $q_0= \pm 1/4$, which can thus
be naturally identified with the {\it elementary charge}. It follows,
therefore, that the {\it Kerr congruence can be equipped with an
electromagnetic field defined directly via its principal spinor and carrying
necessarily a ``unit" electric charge}.

On the other hand, system (\ref{GSE}), though equivalent to (\ref{rbccv}) or (\ref{delx}) and thus to the equations of the shear-free null geodesic congruence (\ref{shearfr}), was originally considered as a particular case of the differentiability conditions in the algebra of complex quaternions, isomorphic to the full algebra of $2\times 2$ complex matrices, $\mathbb{B}\cong Mat(2,\mathbb{C})$ (see e.g.~\cite
{kass7,YadPhys}).
In the present paper, however, we do not intend to discuss the problem of noncommutative analysis, referring the interested reader, e.g. to the profound paper of A.Sudbery~\cite{sud}, to the reviews~\cite{Gsponer}, and to our early contributions on the subject~\cite{kass5,kass1,kass2,kass3}. 

Let us outline the organization of the article. In the second
section we obtain the 4-eikonal equation for the components of the basic
2-spinor, demonstrate the functional dependence of the components of the
twistor associated with the generating system of equations and, thus, obtain the general solution of the latter in the form  of the {\em algebraic} system (\ref{sol}). The third section is devoted to the analysis of the ``restricted'' gauge symmetry of the generating system of equations 
(for which the gauge parameter depends only on the components of the field under transform) 
and its specific relation to the projective transformations in
associated twistor space. In Sec. 4 we study the self-dual structure of
the generating system of equations which follows from its integrability conditions and guarantees that the
source-free Maxwell and Yang-Mills equations hold for the respective fields associated with the solutions of the generating system of equations. We also demonstrate that a 2-spinor composed of the components of the 4-potential matrix $\Phi$ obeys the Weyl equation for spin-1/2 massless fields. In Sec. 5 we rigorously study the relations of the generating system of equations to those of shear-free null geodesic congruences, write out the explicit expressions for the strengths of the associated Maxwell and Yang Mills fields and introduce Riemannian metrics of the Kerr-Schild type. A brief review of the previously obtained ``particle-like" solutions of the generating system of equations and of the associated fields, together with the property of electric charge ``self-quantization", is presented in Sec. 6. In the last section we summarize our findings and conclude the paper.

\section{The Eikonal Equation and General Solution to the Generating System of Equations}
\label{pbccv}

In the matrix representation of the biquaternion algebra $\mathbb{B}$
used in (\ref{GSE}) we regard $\xi$ as a column ($2\times 1$ matrix over
$\mathbb{C}$), $\Phi$ -- as a $2\times 2$ complex matrix of the general type,
and $X$ -- as a hermitian matrix representing the coordinates of Minkowski
space-time. System (\ref{GSE}) is evidently invariant under the global
{\it Lorentz transformations} of coordinates
\begin{equation}\label{symm}
X\rightarrow L X L^+,~~~\xi\rightarrow \bar L^+\xi,~~~\Phi\rightarrow
\bar L^+ \Phi \bar L,
\end{equation}
where $L\in SL(2,\mathbb{C})$ and $\bar L$ is the inverse of $L$. We see that
the field $\xi(x)$ really
transforms as a 2-spinor whereas the components of the field
$\Phi(x)$ constitute a complex 4-vector (later identified with the electromagnetic 4-potential).

In view of the form of transformations (\ref{symm}), $\xi,\Phi$ are
mapped respectively to matrices $\xi_{A^\prime}$ and
$\Phi_{B^\prime A}$, whereas $dX$ -- to a hermitian matrix $d{X^{AA^\prime}}$,
with (un)primed indices $A,...A^\prime,...=0,1;0^\prime,1^\prime$, in the usual
2-spinor notation. In this notation, system (\ref{GSE}) takes the aforementioned form
(\ref{delx}), equivalent to a system of eight partial differential equations

Throughout the paper, we assume that both spinor components $\xi_{A^\prime}$
are nonzero in the considered region of spacetime (otherwise, one can easily show, that the
electromagnetic and other associated fields vanish trivially). We also assume all the functions $\{\xi_{A^\prime}(x),~\Phi_{B^\prime A}(x)\}$ to be analytic
everywhere except at some subset of zero measure where they may be singular.

Some properties of the solution $\xi(x)$ can be inferred directly from
(\ref{delx}). Using the orthogonality identity $\xi^{A^\prime}
\xi_{A^\prime}=0$, one readily finds
\begin{equation}\label{eikxx}
\nabla^{AA^\prime}\xi_{C^\prime} \nabla_{AA^\prime}\xi_{B^\prime}=0,
\end{equation}
which in particular implies the {\it eikonal} equation for any
function $\lambda(\xi_{A^\prime})$ of spinor components
\begin{equation}\label{eikx}
\nabla^{A A^\prime}\lambda \nabla_{A A^\prime}\lambda = 0.
\end{equation}

Returning to system (\ref{delx}) and multiplying it by $\xi^{A^\prime}$
we reduce it back to  system (\ref{rbccv})
where the 4-vector field $\Phi_{B^\prime A}$ has been dispensed with. The latter may be recovered by multiplying (\ref{delx})  by $\iota^{A^\prime}$,
\begin{equation}\label{4pot}
\Phi_{B^\prime A}=\iota^{A^\prime}\nabla_{A A^\prime} \xi_{B^\prime}, 
\end{equation}
with $\iota$  being a 2-spinor conjugate to $\xi$,
\begin{equation}\label{iota}
\xi_{A^\prime} \iota^{A^\prime} =1,
 \end{equation}
defined up to the transformation
\begin{equation}\label{iotagauge}
\iota \mapsto \iota + \alpha \xi,
\end{equation} 
where $\alpha=\alpha(x) \in \mathbb C$ is an arbitrary scalar field. 


Conversely, it follows from (\ref{rbccv}) that the spintensor
$\Phi_{A^\prime A}$ can be defined by (\ref{delx}), implying that (\ref{rbccv}) is equivalent to the
original spinor system (\ref{delx}), hence to the generating system of equations (\ref{GSE}).

We now turn to the solutions of (\ref{rbccv}), or equivalently, of the generating system of equations (\ref{delx}). Since (\ref{sol}) and (\ref{rbccv}) are possibly related, one might expect the latter to possess a twistor structure. To
show this, let us introduce a 2-spinor $\tau^{A}$ related to
$\xi_{A^\prime}$ via the Penrose correspondence (\ref{tv}).
Then the pair of spinors $T = \{T^a\} = \{\xi_{A^\prime},\tau^A\},~
a=0,1,0^\prime,1^\prime$,
constitute a null~\footnote{Hereafter, we shall ignore the imaginary unit $i$ in (\ref{tv});  the condition for a twistor to be null following from (\ref{tv}) takes then the form $\tau^+\xi - \xi^+ \tau =0$.} twistor $T$ {\it incident} with a (real) Minkowski
space-time point represented by $X^{A A^\prime}$.

According to (\ref{tv}), the wedge product of the differentials
$d\xi_{A^\prime}$ and $d\tau^{A}$ is given by
\begin{equation}\label{dtdx}
d \tau^{A} \wedge d \xi_{B^\prime} = X^{AA^\prime}d \xi_{A^\prime} \wedge
d\xi_{B^\prime} +  \xi_{A^\prime}dX^{AA^\prime} \wedge d \xi_{B^\prime} .
\end{equation}
From (\ref{dtdx}) a rather obvious property of the (nontrivial) twistors
immediately follows: {\it at least} a pair of components of the twistor field 
$T^a$ must be functionally independent (with respect to their dependence on
coordinates $X^{AA^\prime}$). Indeed, if we assume that the exterior products $d \tau^{A} \wedge d \xi_{B^\prime}$ and $d \xi_{A^\prime} \wedge
d\xi_{B^\prime}$ all vanish, then by (\ref{dtdx}) we find $ w^{A} \wedge d\xi_{B^\prime}=0$, where the 1-forms $w^{A}=dX^{AA^\prime}\xi_{A^\prime}$ have been introduced. Together with (\ref{GSE}) this implies the condition $\Phi_{B^\prime C} w^{A} \wedge w^{C} =0$, whose solutions are $w^{A} \wedge w^{C}=0$ or $\Phi_{B^\prime C}=0$. It is easy to see that the first solution leads to $\xi_{A^\prime} = 0$, while combining the second solution with (\ref{GSE}) gives $d\xi_{B^\prime}=0$. Excluding the first trivial case, we conclude that $\xi_{A^\prime}$ are constant, and consequently that the two components
of the spinor $\tau^A$ (\ref{tv}) are functionally independent.

If we now subject the spinor $\xi_{A^\prime}$ to the basic system
(\ref{rbccv}), then the other pair of twistor components {\it must depend}
on the first pair. Precisely, we have the following:
\vskip3mm \noindent{\bf Proposition.1}~{\it Iff $\xi_{A^\prime}$ is a
solution of (\ref{rbccv}) then the corresponding twistor $T^a$ has
two and only two functionally independent components}. \vskip1mm

Going to differentials $d\xi_{A^\prime},~d\tau^{A}$ in (\ref{delx}) and in
(\ref{tv}) respectively, we arrive at the following system:
\begin{gather}
d \xi_{B^\prime} = \Phi_{B^\prime A} w^{A},\label{dxw}\\ d\tau^{B} =
X^{BB^\prime} \Phi_{B^\prime A}w^{A}
+w^{B},\label{dtw}
\end{gather}
where as before $w^{A}=dX^{AA^\prime}\xi_{A^\prime}$. Since the differentials of all twistor components are
linear functions of the two 1-forms $w^{A}$ only, it becomes
evident that the exterior product of any three is zero, resulting
in the desired functional dependence.
This functional dependence can be expressed~\cite{kr1} in a more symmetric form (\ref{sol}) by a pair of holomorphic functions $\{\Pi^{(C)}\},~C=0,1$ of four complex variables.

Conversely, the algebraic system (\ref{sol}) implicitly determines
$\xi_{A^\prime}$ and it is easy to check, by differentiating
(\ref{sol}) and multiplying the resulting system by $\xi^{A^\prime}$, that
$\xi_{B^\prime}$ satisfies
\begin{equation}\label{diffpi}
\frac{d
\Pi^{(C)}}{d\xi_{B^\prime}}\xi^{A^\prime}\nabla_{AA^\prime}\xi_{B^\prime}=0,
\end{equation}
which, indeed, is equivalent to (\ref{rbccv}) except at some singular
points (see below). Successively resolving system (\ref{sol}) at each
space-time point $X^{A A^\prime}$ with respect to  $\xi_{A^\prime}$ and
substituting the resulting solution in (\ref{4pot}) to find the corresponding
``potentials" $\Phi_{B^\prime A}$ we obtain a solution to the generating system of equations starting
from the algebraic constraints (\ref{sol}). This furnishes the proof of
proposition.1.

Thus, system (\ref{sol})
implicitly determines the {\it general (analytical) solution} $\{\xi_{A^\prime},
\Phi_{B^\prime A}\}$ of the generating system of equations. Points where equations (\ref{sol}) have
multiple roots, i.e. cannot be {\it uniquely} resolved for
$\xi_{A^\prime}$ according to (\ref{diffpi}) are defined by the condition (compare with
(\ref{sinold}))
\begin{equation}\label{sin}
\det\Vert\frac{d \Pi^{(C)}}{d\xi_{A^\prime}}\Vert =0.
\end{equation}
These points constitute a singular set of the electromagnetic field, which
in the next Section will be associated with $\Phi_{A^\prime A}$.
Together with (\ref{sol}) the last equation allows us to determine the
shape and the time evolution of singularities (see Sec. 6 below).

Geometrically, system (\ref{sol}) defines a 2-dimensional
complex surface in twistor space $\mathbb{C}^4$ (precisely, in the subspace of
null twistors). Points of intersection of this surface
with 2-dimensional planes formed by (null) twistors (\ref{tv}) represent
the solution $\xi_{A^\prime}$ to the generating system of equations ({\it multivalued} in general) for each
fixed space-time point $X^{A A^\prime}$. Singularities are then the pre-images
(in $\bf M$) of the points of twistor space at which the planes (\ref{tv}) are
{\it tangent} to the surfaces (\ref{sol}). Note that we ignore here
the generally considered projective structure of the twistors which,
in the present approach, has some peculiarities and is related to an
unconventional gauge symmetry of the generating system of equations. These issues are discussed in the next section.

In passing, we shall establish an auxiliary result, which will come in handy in the next Section. Rewriting (\ref{dtw}) in partial derivatives we get
\begin{equation}\label{delt}
\nabla_{AA^\prime}\tau^{B} = X^{BB^\prime}
\Phi_{B^\prime A}\xi_{A^\prime} +\xi_{A^\prime}\delta^{B}_{A}~.
\end{equation}
Using the orthogonality $\xi^{A^\prime}\xi_{A^\prime}=0$, we immediately verify
the validity of the following relations:
\begin{gather}
\nabla^{AA^\prime}\xi_{B^\prime}\nabla_{AA^\prime}\tau^{B}=0,\label{eikxt}\\
\nabla^{AA^\prime}\tau^{B}\nabla_{AA^\prime}\tau^{C}=0,\label{eiktt}
\end{gather}
which along with (\ref{eikxx}) lead to the eikonal
equation 
\begin{equation}\label{eikonalt}
\nabla^{AA^\prime} \lambda \nabla_{AA^\prime} \lambda = 0,
\end{equation}
for any function $\lambda(T^a)$ of the four twistor components.

\section{Projective transformations of twistors and the ``weak"
gauge symmetry of the generating system of equations}

For an appropriate electrodynamical interpretation, we need to
establish the gauge invariance of the generating system of equations. This is addressed in this
section. More precisely, we shall study the symmetry of (\ref{delx})
under transformations
\begin{equation}\label{gauge}
\xi_{A^\prime} \to \xi^\prime_{A^\prime}=\alpha(x)\xi_{A^\prime},
\end{equation}
where $\alpha(x)$ is a smooth complex function of coordinates. Using
(\ref{rbccv}), it's readily seen that $\alpha$ cannot be an
arbitrary function of coordinates; it should rather satisfy the constraint
\begin{equation}\label{alph1}
\xi^{A^\prime}\nabla_{AA^\prime}\alpha=0,
\end{equation}
from which, in view of 2-spinors' properties, it follows
\begin{equation}
\nabla_{AA^\prime} \alpha = \rho_A \xi_{A^\prime}
\end{equation}
for some $\rho_A$ and, consequently, the eikonal equation for $\alpha(x)$
\begin{equation}\label{eikalph}
\nabla^{AA^\prime}\alpha \nabla_{AA^\prime}\alpha = 0.
\end{equation}


Together with (\ref{eikonalt}) this leads us to consider the following:
\vskip3mm \noindent{\bf Proposition.2}~{\it Transformations of the
type (\ref{gauge}) are symmetries of (\ref{delx}) iff $\alpha$ is a
function of $T^a$ and $\Phi_{B^\prime A}$ transform according to}
\begin{equation}\label{gradpot}
\Phi_{B^\prime A} \to \Phi^\prime_{B^\prime A} = \Phi_{B^\prime A} +
\nabla_{A B^\prime} \ln \alpha.
\end{equation}

Replacing in (\ref{delx}) $\xi_{A^\prime}$ and $\Phi_{B^\prime A}$ by their
transformed values, after some simplification we obtain the following
condition of the form-invariance of (\ref{delx}):
\begin{equation}\label{alph2}
\xi_{B^\prime}\nabla_{A A^\prime}\alpha - \xi_{A^\prime}\nabla_{A B^\prime}\alpha = 0,
\end{equation}
which is skew-symmetric in $A^\prime, B^\prime$ and therefore equivalent to equation (\ref{alph1}) for $\alpha(x)$.
Assuming now that $\alpha=\alpha(T^a)=\alpha(\xi_{A^\prime},\tau^B)$,
taking its derivatives and using (\ref{delx}) and (\ref{delt}) we prove that
(\ref{alph1}) is identically satisfied, and thus every $\alpha(T^a)$
realizes a symmetry of the generating system of equations. This much takes care of the sufficient part of the
proposition.

To prove the converse, suppose that transformation (\ref{gauge})
is a symmetry of (\ref{delx}). This yields the following:
\begin{equation}\label{dalph}
\xi_{B^\prime} d\alpha = \alpha(\Phi^\prime_{B^\prime A}-
\Phi_{B^\prime A}) w^A,
\end{equation}
where as before $w^{A}=dX^{AA^\prime}\xi_{A^\prime}$. Making use
of (\ref{dxw}) and (\ref{dtw}), it's easy to see that the
exterior product of equation (\ref{dalph}) with any two
differentials of the twistor components vanishes, leading to the
functional dependence of $\alpha$ on the corresponding twistor
components. More symmetrically, this can be expressed as
asserted above, i.e. $\alpha=\alpha(T^a)$.
Passing then to partial derivatives in (\ref{dalph}) we obtain the relation
\begin{equation}\label{delalph}
\xi_{B^\prime}\nabla_{AA^\prime}\ln \alpha = (\Phi^\prime_{B^\prime A}-
\Phi_{B^\prime A})\xi_{A^\prime}~,
\end{equation}
from which the desired transformation rule (\ref{gradpot}) for the ``potentials"
$\Phi_{B^\prime A}$
follows (in a special case when the primed subscripts
are equal). This completes the proof of proposition.2. $\blacksquare$

A few words are in order about the nature of transformations
\begin{equation}\label{rgauge}
\xi \to \xi^{\prime} = \alpha(T^a)\xi,
\end{equation}
which may be called {\it restricted}, or {\it weak} gauge transformations
~\cite{kr2}. Note that according to its definition (\ref{tv}) the conjugate
spinor $\tau^B$ transforms in a similar way
\begin{equation}\label{taugauge}
\tau \to \tau^\prime=\alpha(T^a)\tau
\end{equation}
so that together (\ref{rgauge}) and (\ref{taugauge}) imply that the gauge
symmetry of the generating system of equations may be expressed in twistor form
\begin{equation}\label{tvgauge}
T \to T^\prime = \alpha(T^a) T.
\end{equation}

It's clear that  composition of transformations (\ref{tvgauge}) is a
transformation of the same type. Their associativity is also obvious. But the
existence of inverse transformation is not so evident. However, according to
proposition.1, both $T^a$ and its image $T^{\prime a}$ have only
two functionally independent components, and $\alpha(T^a)$
depends, essentially, on these two components. So we
can always express the two independent components of $T^a$ through
those of $T^{\prime a}$. Substituting them in $\alpha^{-1}(T^a)$ results in
the inverse transformation $T=\alpha^{-1}(T^{\prime a})T^\prime$ which
is of the same type as (\ref{tvgauge}). Hence {\it these
transformations constitute a group}! In fact it is a proper
subgroup of the full $\mathbb{C}(1)$-gauge group of transformations
(\ref{gauge}), the latter itself not being generally a symmetry of the generating system of equations.
The statement that this subgroup is proper, becomes quite evident if we recall that
$\alpha(T^a)$ necessarily obeys the eikonal equation (\ref{eikalph}).

Finally, we note that under transformations (\ref{rgauge}) the {\it trace}
of the matrix 1-form $A:= Tr(\Phi dX) = \Phi_{A^\prime A} dX^{A A^\prime}$
transforms gradient-wise
\begin{equation}\label{potgauge}
A \to A + d\ln \alpha,
\end{equation}
just as the electromagnetic potential 1-form does under ordinary gauge transformations. Together with the
4-vector properties of $\Phi$ under the Lorentz transformations (\ref{symm}) this leads us to adopt the  interpretation of the 1-form $A$ and the corresponding gauge-invariant 2-form $F=dA$ as the electromagnetic 4-potential and the electromagnetic field strengths respectively
(of course, up to an arbitrary scale factor). In the following Section, we further elaborate on this electromagnetic interpretation by 
obtaining Maxwell equations for the 2-form $F$.

Let us now look at the gauge transformations (\ref{tvgauge}) from the
standpoint of the geometry of twistor space. The Abelian nature of these transformations
results in the {\it ratio} of any two twistor components
$T^a$ remaining invariant, thus hinting at their {\it projective}
origin. Not only the planes (\ref{tv}), but also the surfaces (\ref{sol}) are
form-invariant under transformations (\ref{tvgauge}) and, consequently,
give rise to  another solution of the generating system of equations (with the same electromagnetic 2-form
$F$). So we may consider the {\it equivalence classes} of the solutions (and of
the surfaces (\ref{sol}) respectively) which can be
obtained one from another via the gauge transformations (\ref{tvgauge}).
That's why, we may restrict ourselves to consider only
the {\it projective twistor space} $\mathbb{C}P^3$. However, a projective structure
of this type differs essentially from the conventional one, which originates
from the transformations of the full gauge group (\ref{gauge}). We shall return to this issue in Sec. 5. For the time being we shall deal with the structure of the full space of (null) twistors.

\section{Self-duality, source-free gauge fields and Weyl equation}
\label{int}
As mentioned in the introduction, from a geometric point of view, the generating system of equations (\ref{GSE}) is
the condition which must be met for a spinor $\xi(x)$ to be covariantly
constant with respect to the matrix-connection 1-form
\begin{equation}\label{conn}
\Omega=\Phi dX .
\end{equation}
In view of (\ref{conn}), the original generating system of equations (\ref{GSE}) may be
written as follows:
\begin{equation}\label{ccv}
d \xi=\Omega \xi.
\end{equation}

Before we proceed further, it is worth noting that in a 4-vector representation the generating system of equations is equivalent to a system of equations defining a {\em covariantly constant vector field} w.r.t. a specific affine connection.  
Its particular form follows from a {\it relativistic version of biquaternion algebra} introduced orginally in~\cite{Grgin} and is considered in~\cite{Mark16}. It is, however, worth noting that equations for covariantly constant fields on the background of Weyl geometry~\cite{kr5} or a geometry with torsion determined by its trace~\cite{KasJos15} can also be used for a transparent geometric treatment of electromagnetism and possess a number of properties closely related to those of the generating system of equations.  In particular, any covariantly constant vector field in ordinary Weyl space is, remarkably, a shear-free null geodesic congruence on the underlying Minkowski background.
  


Reverting to the generating system of equations, we recall that it is an {\it overdetermined system} (comprising eight equations for six unknown functions), from which both the spinor and the ``connection" gauge fields 
are to be determined. The dynamics of the connection field $\Omega(x)$ can
be obtained by external differentiation of (\ref{ccv}), which yields
\begin{equation}\label{compat}
R \xi \equiv (d \Omega - \Omega \wedge \Omega) \xi = 0,
\end{equation}
where in parentheses is the matrix {\it curvature 2-form} $R$ of the connection
(\ref{conn}). Since the spinor $\xi$ is not arbitrary, but subject to
(\ref{ccv}), the {\it integrability conditions} (\ref{compat}) don't imply
the triviality of curvature, 
instead they lead to its {\it self-duality}~\cite{kass1,kass2}.

To demonstrate this, we note that for connection ({\ref{conn}),
the curvature $R$ is of the following, rather specific form
\begin{equation}\label{curv}
R = (d \Phi - \Phi dX \Phi) \wedge dX \equiv \pi \wedge dX ,
\end{equation}
where a new matrix-valued 1-form $\pi$ emerges, with the components
\begin{equation}\label{compon}
\pi_{A^\prime C}=\pi_{A^\prime C B B^\prime}dX^{BB^\prime}=
(\nabla_{BB^\prime}\Phi_{A^\prime C}-\Phi_{A^\prime B}\Phi_{B^\prime C})
dX^{BB^\prime}.
\end{equation}
In terms of $\pi$, the integrability conditions (\ref{compat}) read $(\pi \wedge dX) \xi =0$, or in spinor notation
\begin{equation}
\pi_{A^\prime C B B^\prime} dX^{B B^\prime} \wedge dX^{C C^\prime}
\xi_{C^\prime} =0 .
\end{equation}
Making use of symmetry properties, we obtain from the last relation
\begin{equation}
\pi_{A^\prime ~ C(B^\prime}^{~~C}\xi_{C^\prime)} = 0,
\end{equation}
so that for any nontrivial solution $\xi(x)$ it follows
\begin{equation}\label{compi}
\pi_{A^\prime~CB^\prime}^{~~C}\equiv \nabla_{C B^\prime}\Phi_{A^\prime}^{~~C}+
\Phi_{B^\prime C}\Phi_{A^\prime}^{~~C} = 0.
\end{equation}

Decomposing, in the usual way, the curvature (\ref{curv}) into the self- and
antiself-dual parts it is easy to verify that equations (\ref{compi}) are
just the conditions for its self-dual part to vanish, whereas the other
(antiself-dual) part $\bar R$ takes the form
\begin{equation}\label{antiself}
\bar R_{A^\prime(BC)}^{~~C^\prime}=\nabla_{~(B}^{C^\prime}\Phi_{|A^\prime| C)} -
\Phi_{~(B}^{C^\prime}\Phi_{|A^\prime| C)}
\end{equation}
and satisfies additional integrability conditions $\bar R \xi=0$ (we'll not make
use of them below).

Thus, though the curvature 2-form (\ref{curv}) of the connection 1-form
(\ref{conn}) is not identically (anti)self-dual (that is, (anti)self-dual in the ``strong" sense),
it necessarily becomes (anti)self-dual {\it on the solutions of the generating system of equations}. For
this reason, the present property has been called {\it weak self-duality}~\cite{kr1}.

Physically, expression (\ref{antiself}) represents the strength of
a matrix gauge field. In particular, its trace $F_{(BC)}=\bar
R_{A^\prime~~(BC)}^{~~A^\prime}=\nabla_{~(B}^{A^\prime}\Phi_{|A^\prime|C)}$
corresponds to the aforementioned electromagnetic field strength $F=dA$, whereas
the trace-free part of (\ref{antiself}) defines the strength of a {\it Yang-Mills
type' field}~\footnote{Owing to the restricted (weak) gauge symmetry this
is not, precisely, what is usually regarded as, say, the $SL(2,\mathbb{C})$
Yang-Mills fields; nonetheless, these equations are identical in {\it form}, and the restrictions are imposed merely {\it on the set of solutions admissible by the generating system of equations}.}.

Indeed, in view of the {\it Bianchi identities}
\begin{equation}\label{bianchi}
dR \equiv \Omega \wedge R - R \wedge \Omega,
\end{equation}
self-duality of curvature $R+iR^* = 0$ leads to the source-free Maxwell
equations for the electromagnetic 2-form $F=Tr(R)=R_{A^\prime}^{~~A^\prime}$
\begin{equation}\label{maxeq}
dF^*=0= dF \equiv 0,
\end{equation}
and to equations of the Yang-Mills type~\cite{kass2} 
\begin{equation}\label{YMeq}
d{\bf F}^* - [{\bf \Omega},{\bf F}^*] =0, 
\end{equation}
for the trace-free part of the curvature 
\begin{equation}\label{YM}
{\bf F}_{A^\prime}^{~~B^\prime}=R_{A^\prime}^{~~B^\prime}-\frac{1}{2}
F\delta_{A^\prime}^{~B^\prime}, 
\end{equation}
corresponding to the trace-free part $\bf \Omega$ of the connection 1-form (\ref{conn}).

Generally speaking, the electromagnetic 2-form $F$ is a $\mathbb{C}$-valued
field, yet in view of its self-duality it reduces to an
$\mathbb{R}$-valued 2-form $\tt{F}$ related to $F$ through
\begin{equation}\label{freal}
F=\texttt{F}- i \texttt{F}^*,
\end{equation}
for which Maxwell equations in free space, in view of their linearity, hold too, so that the number of its
degrees of freedom is just that of the ordinary electromagnetic field.

Explicitly, from the symmetric part of the integrability conditions (\ref{compi}) we get
\begin{equation}\label{eh}
\nabla_{C(B^\prime} \Phi_{A^\prime)}^{~~C} = 0 ~~~\Longleftrightarrow ~~~ F + iF^* = 0.
\end{equation}
{\it The complex self-duality conditions (\ref{eh}) are (locally) completely equivalent to the real vacuum Maxwell equations}~\cite{kass4} so that all the solutions of the latter can be obtained from them and vice versa. However, not every solution of (\ref{eh}) can be augmented to a solution of the overdetermined generating system of equations itself. Such a state of affairs leads, in particular, to the fundamental property of charge quantization (see Sec.~\ref{bound} below). 

Note also that from the
skew-symmetric part of (\ref{compi}) follows an
additional ``inhomogeneous Lorentz condition"~\cite{kass1,kass2} for the
$\mathbb{C}$-valued electromagnetic potentials $\Phi_{A^\prime A}~= A_\mu$:
\begin{equation}\label{loren}
\nabla_{CC^\prime} \Phi^{C^\prime C} + \Phi_{C^\prime C}\Phi^{C^\prime C} = 0
~~~\Longleftrightarrow ~~~\partial_\mu A^\mu + 2 A_\mu A^\mu = 0,
\end{equation}
which should also hold identically on every solution of the generating system of equations. Condition
(\ref{loren}) is by no means gauge invariant in the ordinary sense, but it {\it is}
invariant w.r.t. the weak gauge transformations (\ref{rgauge}),
provided the potentials together with the principal spinor satisfy the generating system of equations.

Yet another remarkable constraint follows from the structure of the self-duality conditions (\ref{eh}) together with the additional condition (\ref{loren}). Indeed, making use of the weak gauge transformations 
(\ref{rgauge}),(\ref{gradpot}), one can get rid of a pair of components of the 4-potential matrix~ $\Phi_{C^\prime C}$ (this will be demonstrated explicitly in Sec.~5), in such a way that the norm $\Phi_{C^\prime C}\Phi^{C^\prime C}$ in ({\ref{loren}) vanishes. The joint system of equations (\ref{eh}) and (\ref{loren}) then yields
\begin{equation}\label{weyl}
 \nabla_{AB^\prime} \phi^{A} = 0, 
 \end{equation}
 where $\phi(x)=\{\phi^A\}$ stands for a 2-spinor field composed of two {\it nonvanishing} components of the spintensor $\Phi_{A^\prime A}$. Equation (\ref{weyl}) evidently reproduces the {\it Weyl equation} in its canonical form. Thus, {\it for any solution of the generating system of equations, the components of the 4-potential matrix form a 2-spinor field which necessarily satisfies the Weyl equation for a spin-1/2 massless field}. 
    
As to the Yang-Mills type fields ${\bf F}_{A^\prime}^{~~B^\prime}$, they may always be expressed via the
electromagnetic field strengths and the ratio of the spinor components (see Sec.~\ref{grav} below). Note also that neither the real nor the
imaginary part of the trace-free curvature ${\bf F}_{A^\prime}^{~~B^\prime}$ taken separately satisfy the source-free YM equations, in view
of the  nonlinearity of the latter. Therefore, the Yang-Mills-like fields here {\it are
necessarily complex}. Further details on the properties of Yang-Mills fields in the present approach can be found in~\cite{kass2}.

\section{The generating system of equations and projective structure of null shear-free congruences}
\label{grav} 

Recall that via the elimination of potentials $\Phi_
{A^\prime A}$, the generating system of equations (\ref{GSE}) takes the form (\ref{rbccv}). In this
form, the generating system of equations is, in fact, equivalent to the canonical equations defining {\it shear-free null
geodesic congruences} so that every, suitably rescaled, shear-free null geodesic congruence
gives a 2-spinor solution of the generating system of equations.

Indeed, once the solution $\xi_{A^\prime}(x)$ to (\ref{rbccv}) is found, then
a field of a null 4-vector $k_\mu(x),~ k_\mu k^\mu =0$, can be defined as
\begin{equation}\label{cong}
k=k_\mu dx^\mu = \xi_A\xi_{A^\prime} dX^{AA^\prime}.
\end{equation}
Its field lines define a null congruence for which the
shear-free criterion~\cite{pr}
\begin{equation}\label{sfc}
\xi^{A^\prime} \xi^{B^\prime} \nabla_{AA^\prime} \xi_{B^\prime}=0
\end{equation}
follows readily from (\ref{rbccv}). Hence, every solution $\xi_{A^\prime}$ of
the generating system of equations actually defines a shear-free null geodesic congruence. Contrary to
(\ref{rbccv}), the equations of a shear-free null geodesic congruence (\ref{sfc}) are invariant
under the full complex Abelian group of rescaling (\ref{gauge}) and reduce to
the system of two equations in partial derivatives~\cite{pr}
\begin{equation}\label{gsol}
\nabla_w G=G\nabla_{u}G,~~~\nabla_{v}G=G\nabla_{\bar w}G,
\end{equation}
where $G=\xi_{1^\prime}/\xi_{0^\prime}$ is the gauge invariant,
and the following generally accepted notation has been used:
\begin{equation}\label{zcoor}
X^{AA^\prime}=
  \begin{pmatrix}
    u & w \\
    \bar w & v
  \end{pmatrix}\equiv
  \begin{pmatrix}
    x^0+x^3  & x^1-ix^2 \\
    x^1+ix^2 & x^0-x^3
  \end{pmatrix}
\end{equation}
The four real quantities $\{x^i; x^0\},~i=1,2,3$ correspond to the Cartesian space and time
coordinates, respectively. Note that the individual spinor components
$\xi_{A^\prime}$ remain indeterminate by the equations of a shear-free null geodesic congruence which
impose restrictions only on their quotient $G(x)$.

Let us compare this with the generating system of equations (\ref{rbccv}). The latter is equivalent
to a system of {\it four} equations for only two spinor components
$\xi_{A^\prime}$
\begin{equation}\label{cbccv}
\nabla_{w}\xi_{A^\prime}=G\nabla_u \xi_{A^\prime}, ~~~
\nabla_v \xi_{A^\prime} = G\nabla_{\bar w} \xi_{A^\prime}
\end{equation}
from which, of course, (\ref{gsol}) follow for the quotient $G(x)$.
Multiple solutions of (\ref{cbccv}) with the same $G$ correspond to different
potentials but have the same strength of the associated electromagnetic field.
In view of this, from now on, we identify the generating system of equations with those of a shear-free null geodesic congruence by considering only the {\it projectively (gauge) invariant part} of the generating system of equations
represented by system (\ref{gsol}) (one may regard this as fixing the gauge to
$\xi_{0^\prime}=1$).
In particular, the gauge-invariant {\it strength} of the
electromagnetic field should depend only on the quotient $G(x)$ and,
therefore, can be defined for a generic shear-free null geodesic congruence (see (\ref{streng2}) below).


The general analytical solution of (\ref{gsol}) for $G(x)$
immediately follows from Proposition.1 in the form of an algebraic equation
\begin{equation}\label{rsol}
\Pi(G, \tau^0, \tau^1)= \Pi(G,u+w G, \bar w+vG)=0,
\end{equation}
which implicitly determines the function $G(x)$. Here, $\Pi$ is an
arbitrary holomorphic function of three complex variables. Equation (\ref{rsol})
expresses the functional dependence of the three components
$G,\tau^0,\tau^1$ of the projective twistor $T^a$ associated with the
solutions of the generating system of equations. For the shear-free null geodesic congruence the equivalent result is the
well known {\it Kerr theorem}~\cite{pr}. Note that the solutions of (\ref{gsol})
in the form (\ref{rsol}) are defined everywhere except at the points of the singular set (\ref{sin})
whose equation now reduces to
\begin{equation}\label{rsin}
P:=\frac{d\Pi}{dG}=0.
\end{equation}

By multiplying the two equations of (\ref{gsol}), we obtain once more the
4-eikonal equation for $G(x)$ in the form
\begin{equation}\label{wafront}
\nabla_u G\nabla_v G -\nabla_w G \nabla_{\bar w} G = 0,
\end{equation}
while by differentiating them, we verify that $G(x)$ {\it satisfies also the linear
d'Alembert equation}~\cite{Trish,kr1} (see also~\cite{kw})
\begin{equation}\label{wave}
\Box G(x) \equiv (\nabla_u\nabla_v - \nabla_w\nabla_{\bar w})G(x) = 0.
\end{equation}
Note that in view of (\ref{wafront}) every $C^2$-function $\lambda(G)$
is also harmonic on the solutions of the generating system of equations,
\begin{equation}\label{harm}
\Box \lambda(G) = 0.
\end{equation}

Using now expression (\ref{4pot}) for the potentials $\Phi_{A^\prime A}$
and taking into account (\ref{wafront}), one can
express the electromagnetic field strengths via the 2-nd order derivatives
of $\Lambda:= \ln G$ as
\begin{equation}\label{streng}
F_{00}=\nabla_u\nabla_{\bar w}\Lambda,~~F_{11}=\nabla_v\nabla_w\Lambda,~~
F_{01}=\nabla_w\nabla_{\bar w}\Lambda ,
\end{equation}
so that the source-free Maxwell equations are guaranteed to hold 
for (\ref{streng}) in view of the wave equation (\ref{harm}) for
$\Lambda=\lambda(G)$. Twice differentiating the identity (\ref{rsol}) w.r.t.
the space-time coordinates, we finally obtain for the strengths (\ref{streng})
the following symmetric expression:
\begin{equation}\label{streng2}
\boxed{F_{AB}=\frac{1}{P}\left(\Pi_{AB}-\frac{d}{dG}\left(\frac{\Pi_A\Pi_B}{P}\right)\right)}
\end{equation}
where the function $P$ is defined by (\ref{rsin}) and $\{\Pi_A,\Pi_{AB}\},~
A,B=0,1$ denote the (1-st and 2-nd order) partial derivatives of $\Pi$ w.r.t. its twistor arguments $\tau^0,\tau^1$. 

The strengths of the triplet of the Yang-Mills field defined by the 2-form (\ref{YM}) can now be represented in explicit form~\cite{diss} 
\begin{equation}\label{YMG}
{\bf F} = \{-GF, iGF, -F\}, 
\end{equation}
its modulus (in isotopic 3-space) being equal to that of the Maxwell field, ${\bf F}^2 = F^2$.

It is worth noting that in the above gauge $\xi_{0^\prime}=1$, as it follows from (\ref{4pot}),  the two components $\Phi_{0^\prime A}$ vanish, leaving $\phi_A:=\Phi_{1^\prime A} =\nabla_{A 1^\prime} \ln G$. Upon differentiating these expressions, in view of the d'Alembert equation (\ref{harm}), one obtains
\begin{equation}\label{weylproj}
\nabla_{\bar w} \phi_0 = \nabla_u \phi_1, ~~~\nabla_v \phi_0 = \nabla_w \phi_1 .
\end{equation}
These are precisely the Weyl equations (\ref{weyl}) for the 2-spinor field $\phi(x)$, which in the considered gauge are equivalent to the self-duality conditions (\ref{eh}) supplemented with the ``inhomogeneous Lorentz condition" (\ref{loren}). 

Finally, the intimate relation between the generating system of equations and the shear-free null geodesic congruence makes it possible to introduce one more  geometrophysical structure -- an effective {\it Riemannian metric}. In fact, 
it's well-known~\cite{dks,ksmh} that the deformation of the flat Minkowski metric
$\eta_{\mu\nu}$ into a metric $g_{\mu\nu}$ of the Kerr-Schild type
\begin{equation}\label{metr}
g_{\mu\nu} = \eta_{\mu\nu} + hk_\mu k_\nu,
\end{equation}
preserves the main characteristics of a shear-free null geodesic congruence (geodesity, twist and
shear). Here $h$ is a scalar field
and the congruence $k$ given by (\ref{cong}) takes a projectively invariant form
\begin{equation}\label{cong2}
k=du+\bar G dw + G d\bar w + G\bar G dv ,
\end{equation}
$\bar G$ being the complex conjugate of $G$.
We now make use of the results of the classical paper~\cite{dks}, where it was proved that the metric (\ref{metr}) satisfies  Einstein-Maxwell
electrovacuum system for any function $G$ obeying the algebraic constraint
(\ref{rsol}), with a function $\Pi$ {\it linear} in twistor arguments
$\tau^0,\tau^1$~:
\begin{equation}\label{PiG}
\Pi = \varphi + (qG + s)\tau^1-(pG + \bar q)\tau^0.
\end{equation}
Here, $\varphi=\varphi(G)$ is an arbitrary analytic function of
the complex variable $G$, $s$ and $p$ are real constants and $q$
is a complex constant. Without going into the details, we note
that according to the results of~\cite{dks}, the scalar field $h$
in (\ref{metr}) is determined, up to an arbitrary real
constant, by the function $\Pi$ and another function $\Psi(G)$
independent of $\varphi(G)$ and related to the electromagnetic
field arising therein. These electromagnetic fields satisfy Maxwell
equations in the curved space-time with metric (\ref{metr}), and generally
are distinct from those emerging in our approach and defined
in the {\it flat} space-time \footnote{At the same time  both these fields
are generally different from the fields which may be defined for shear-free null geodesic congruence
using the Penrose twistor transform ~\cite{pr,prmc} as well as from the null fields constructed via the Robinson
procedure~\cite{rob}}. However, {\it for the most fundamental Reisner-N\"ordstrem
 and Kerr-Newman solutions these fields coincide}~\cite{kr2} the only (but important!) difference being in
 that in our approach the electric charge is fixed in magnitude by the generating system of equations itself
 (see the next Sec.~\ref{bound}).

Moreover, it was shown in~\cite{dks,kw,burin} that the singularities of the Riemann curvature of
the Kerr-Schild metric (\ref{metr}) are given by (\ref{rsin}), which, according to (\ref{streng2}), also determines the singular locus of the electromagnetic
field~\footnote{The associated Yang-Mills field (\ref{YMG}) possesses additional string-like singularities defined by those of the function $G$ itself and resembling the well-known Dirac string. The Weyl 2-spinor field (\ref{weyl}) defined by the 4-potential matrix exhibits singularities of similar string-like type}.    

  Hence, with every solution of the generating system of equations an electromagnetic,
a $\mathbb{C}$-valued Yang-Mills and a Kerr-Schild metric (precisely, curvature) fields can be naturally associated, their principal singularities being determined by (\ref{rsin}) and therefore coincide in space and time giving rise to {\it a unique singular source for a number of
fundamental physical fields} with well-defined shape and dynamics.
This makes it possible, in the framework of
the classical field model based on the generating system of equations, {\it to consider ``particle-like"
field distributions represented by common singularities of all the fields involved}. 
This is illustrated on concrete examples in the next section.

\section{Electric Charge and  Particle-Like Solutions of
the Generating System of Equations}
\label{bound} In this Section we present a few solutions of the generating system of equations
(and thus of the equations of a shear-free null geodesic congruence). They all may
be obtained by an appropriate choice of the function $\Pi$
and the subsequent resolution of the algebraic constraint (\ref{rsol}). To keep things as simple as possible we confine ourselves to functions $\Pi$ {\it quadratic} in G; more complicated examples are presented, e.g. in~\cite{kass4}~\footnote{Solutions obtained from linear twistor functions result in zero electromagnetic and Yang-Mills field strengths (\ref{streng},\ref{YMG})}.

The fundamental {\it static} solution is generated by the following function:
$$
\Pi = G\tau^0-\tau^1 + 2ia \equiv G(u+ w G)-(\bar w + vG)+ 2ia=0,
$$
($a=Const \in \mathbb{R}$), from which we get
\begin{equation}\label{kerr}
G = \frac{\bar w}{(z+ia)\pm r_*}\equiv\frac{x+iy}{(z+ia)\pm \sqrt{x^2+y^2+(z+ia)^2}}.
\end{equation}
The EM field (\ref{streng}) associated with this solution takes
the form ~\cite{kass1,kr2}
\begin{equation}\label{elctrfd}
\vec E - i\vec H = \pm\frac{\vec r_*}{4(r_*)^{3/2}};~~~~~~~~(\vec E + i\vec H = 0),
\end{equation}
where $\vec r_* =\{x,y,z+ia\}$, has a {\it ring singularity} of radius $a$, with an
electric charge $q=\pm 1/4$ (in the dimensionless units we use), and
a magnetic dipole and an electric quadrupole moments equal to $qa$ and $qa^2$,
respectively~\cite{lopes}. Apart from the restriction on charge,
the electromagnetic field (\ref{elctrfd}) and the Riemannian metric defined
by (\ref{metr}) via the shear-free null geodesic congruence (\ref{cong2}), under the appropriate choice of the
scalar function $h(x)$, reproduce exactly the EM field and metric of the Kerr-Newman solution (in the
coordinates used in~\cite{dks}). Particularly, for $a=0$ the solution (\ref{kerr})
represents the {\it stereographic map} $S^2\to \mathbb{C}$, while the associated fields
are the Coulomb field and the Reisner-N\"ordstrem metric respectively.


The self-duality
condition (\ref{eh}) together with the gauge symmetry of the generating system of equations ensures the
relation $q=N/4, N\in \mathbb{Z}$ for
the values of electric charge associated with every solution of the generating system of equations.
This property has both topological and dynamical origins, the latter
being related to the overdetermined structure of the generating system of equations. The proof of the
general theorem on integer-valued charge can be found in ~\cite{kass4}. Unlike the
purely topological approaches~\cite {ranada2,zhuravl} to the problem of
quantization of electric charge, in the present
framework, the charge of the fundamental static solution (\ref{kerr})
can take only one fixed value and can thus be naturally identified with the
{\it elementary charge}. Together with the well-known property of the Kerr-Newman
solution to fix the gyromagnetic ratio $g=2$ (equal to that
of the Dirac particle~\cite{carter}), the emergence of an elementary electric
charge within the theory makes it more legitimate to interpret the
fundamental solution (\ref{kerr}) as a classical model of the electron (in comparison, say, with the models of Lopez~\cite{lopes}, Israel~\cite{israel}, Burinskii~\cite{burin1,BURElectron} and Newman~\cite{NEWMANElectron} based on Einstein-Maxwell theory itself).

According to a general theorem proved in~\cite{kw} the {\it static} solutions
to the shear-free null geodesic congruence equations (and thus to the generating system of equations) with a bounded in 3-space singular set are exhausted by the Kerr solution (\ref{kerr}) (up to 3-translations and
3-rotations). If, however, we relax the static condition and look outside the class of functions (\ref{PiG}) dealt with in~\cite{dks}, we
discover a lot of time-dependent solutions with bounded
singularities of different dimensions, 3-shapes and time evolutions. Such solutions with bounded structure of the common singular locus may be called 
{\it particle-like}~\cite{kt}. They constitute a wide and physically
interesting class of solutions of the equations of shear-free null geodesic congruence (the generating system of equations) and of the associated field equations, in particular Maxwell source-free equations. Some of these
solutions seem to be quite unfamiliar in classical electrodynamics. 

As an example, consider the {\it axisymmetric} solution of the particle-like type generated
by the function~\cite{kr1,diss}
\begin{equation}\label{joe}
\Pi = \tau^0\tau^1 + b^2 G^2 = 0, ~~~b=Const.
\end{equation}
For real $b$, it admits two singularities, with necessarily ``elementary" charges $+1/4$ and
$-1/4$, undergoing head-on hyperbolic motion. The corresponding electromagnetic field
\begin{equation}\label{fieldborn}
E_\rho=\pm\frac{8b^2\rho z}{\Delta^{3/2}},~~~E_z = \mp\frac{4b^2 M}{\Delta^{3/2}},~~~
H_\varphi =\pm\frac{8b^2\rho t}{\Delta^{3/2}},
\end{equation}
is the well-known {\it Born solution}~\cite{ror}. Here the
following notation is used:
$$\rho^2 = x^2+y^2,~~~s^2 = t^2-z^2,~~~M=s^2+\rho^2+b^2,~~~\Delta=M^2-4s^2\rho^2,$$
and the singularities are defined by the condition $\Delta=0$. However, for a general complex $b = b_0+ib_1$, with real $b_0$ and $b_1$, the singular set consists of two Kerr-like rings (see Fig.~\ref{pic1}), each of radius $|b_1|$, carrying opposite ``elementary'' charges and executing hyperbolic motion along the z-axis: $z=\pm\sqrt{b_0^2+t^2}$. This, to the best of our knowledge, previously unknown distribution of electromagnetic field could be interesting in the general context of the Einstein-Maxwell theory. Moreover, for imaginary $b=ia,~ a\in \mathbb{R}$ one obtains another, {\it  electrically neutral solution} with a ring-like singularity at $t=0$ which then expands into a {\it torus}. At times $t>\vert a \vert$ the singularity turns into a {\it self-intersecting torus}, as depicted in Fig.~\ref{pic2}.

\begin{center}
\begin{figure}[ht]
\centering
\includegraphics[angle=0,scale=0.7]{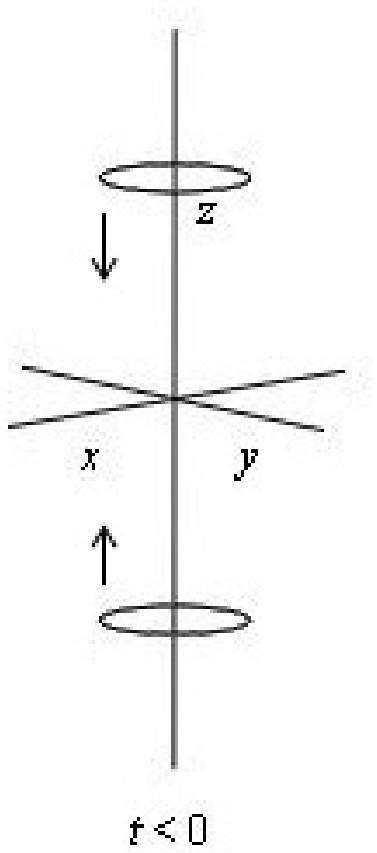}
\includegraphics[angle=0,scale=0.7]{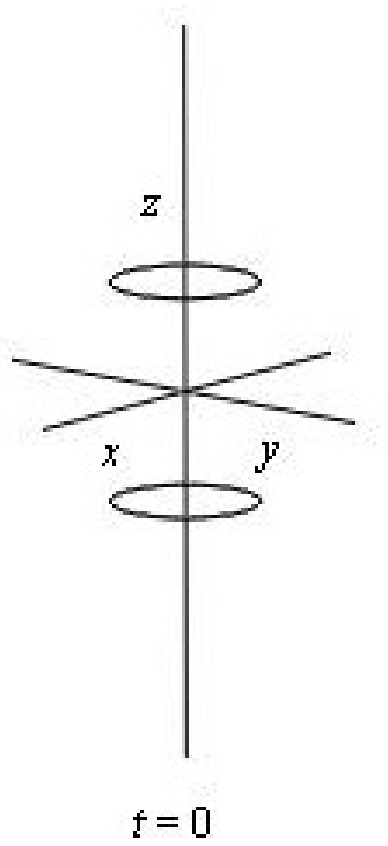}
\includegraphics[angle=0,scale=0.7]{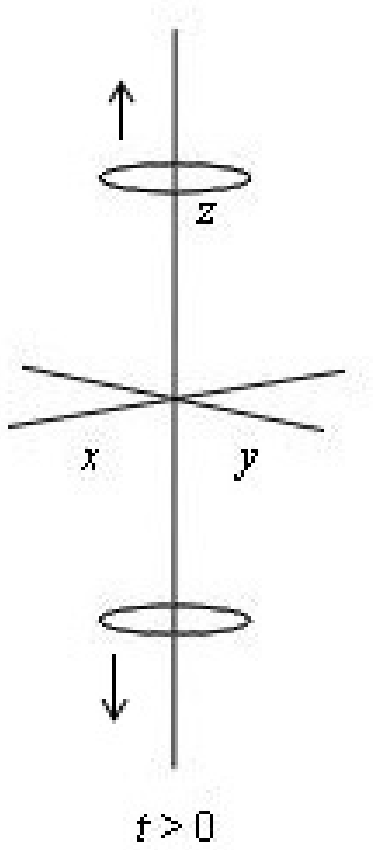}
\vskip 1.7 cm
\caption{The double Kerr-like ring as the singular set of the EM field (\ref{fieldborn}) with complex parameter $b= b_0+ib_1$.}
\label{pic1}
\end{figure}
\end{center} 

Other particle-like solutions for which the singularity has a plane
{\it 8-figure} shape at $t=0$, as well as a wave-type solution with a
{\it helix-like} singularity were also presented in~\cite{kt}. The latter is the counterpart of the
electromagnetic wave in the present approach. A more complicated example simulating the process of pair annihilation can be found in~\cite{kass4}. 

It should be emphasized however that the singularities defined by the condition (\ref{rsin}) are, in general, {\it string-like}, i.e. form a set of 1-dimensional curves moving in 3-space. Indeed, this $\mathbb C$-condition reduces to 2 real equations for 4 coordinates on $\bf M$. Some properties of the arising string structures are presented and explored in~\cite{burin2}.	      

\begin{center}
\begin{figure}[ht]
\includegraphics[angle=-90,scale=0.35]{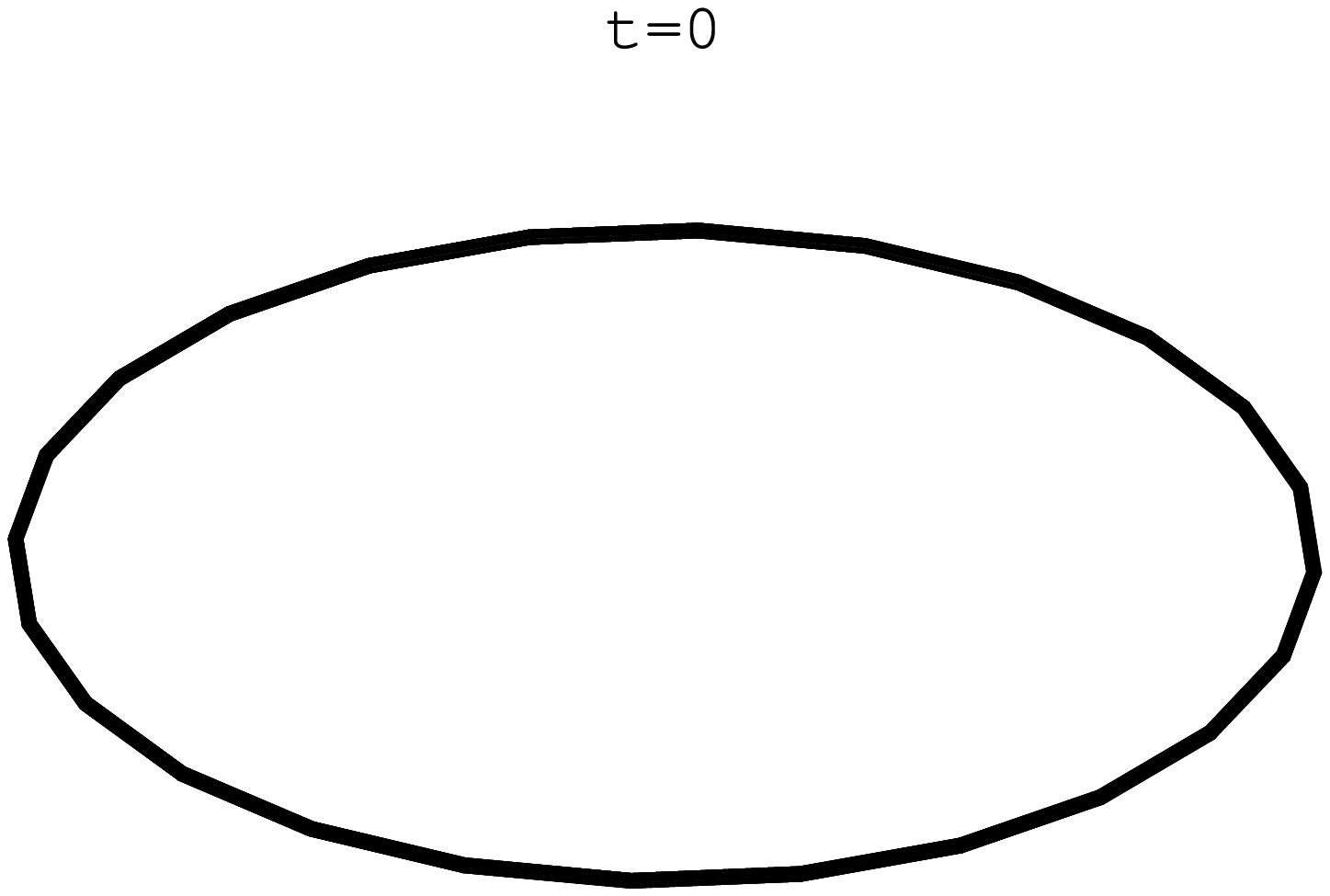}
\includegraphics[angle=-90,scale=0.35]{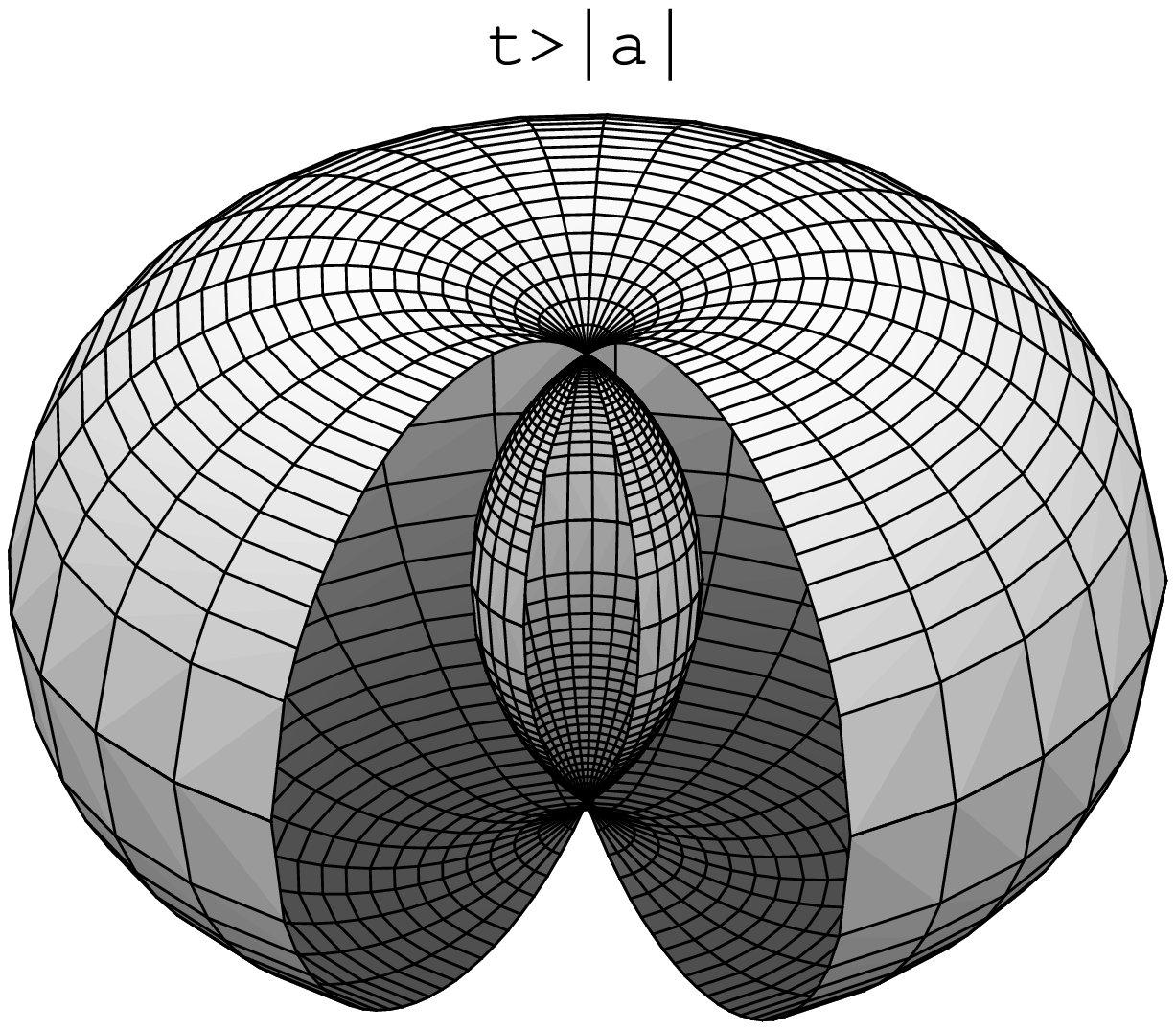}
\vskip 1.7 cm
\caption{The singular set of the electromagnetic field (\ref{fieldborn}) arising from (\ref{joe}), at the initial ($t=0$) and later
($t>\vert a \vert$) times.}
\label{pic2}
\end{figure}
\end{center}


We see that a number of exact solutions to the source-free Maxwell equations
can be obtained in a purely algebraic way, some of which are previously unknown. These solutions are defined everywhere except at the points where the electromagnetic field blows up to infinity. These points constitute a locus which may be 0-, 1- or 2-dimensional
and for a ``particle-like" solution is bounded in 3-space. In general, it is not
possible to cover such a set with a $\delta$-like source (due to the
multivaluedness of the Kerr-type solutions). Nonetheless, the 3-shape and time evolution of the singularities are well
defined and nontrivial due to the {\it "hidden nonlinearity"}~\cite{ranada2} of
Maxwell equations in this theory, ``inherited" from the underlying generating system of equations (the equations of shear-free null geodesic congruence). While the latter ensures the existence of some ``selection rules" for the solutions of Maxwell equation compatible with the spinor structure of the generating system of equations, primarily for the electric charge, it also leads to the violation of the superposition principle (superposed solutions satisfy Maxwell equations but not necessarily the generating system of equations itself). A detailed discussion of the status of singular particle-like solutions may be found in~\cite{kt}. 

\section{Conclusion}
\label{con}
Shear-free congruences of rectilinear light-like rays on the Minkowski space-time $\bf M$ (or its Kerr-Schild deformations) seem to be one of the simplest geometrical objects. Nonetheless, they turn out to be deeply related to a lot of other fundamental geometrical and physical structures. Some of them, in particular twistor geometry, have been known for a long time, as famously expressed by the Kerr theorem (\ref{kerr}). 

However, the situation is even more remarkable. The shear-free null geodesic congruence equations turn out to be
closely related to the form of the generating system of equations (\ref{GSE}) which explicitly exhibits a number of fundamental symmetries and connections with relativistic fields. Specifically, the complex eikonal, wave, and Weyl equations as well as $SL(2,\mathbb C)$ Yang-Mills and Maxwell source-free equations hold identically on the solutions of generating system of equations (the equations of shear-free null geodesic congruence). 

On the other hand, the generating system of equations {\it select} from the solutions of relativistic field equations a peculiar and physically appropriate class. Together with the nontrivial time-dependent structure of common (bounded and carrying integer-valued electric charge) singularities of these fields, we arrive at a self-consistent Lorentz invariant dynamics of a set of such ``particle-like'' formations which is purely algebraic/geometric in origin. Making use of the asymptotic behavior of Kerr-Shild type solutions of Einstein equations explicitly defined by the generating system of equations, one could naturally endow each of the particle-like formations with spin and mass.

Moreover, such collective dynamics appear to be completely conservative. Indeed, consider 
a shear-free null geodesic congruence generated by a point-like source moving arbitrary along a curve in $\bf M$ (or its complex extension~\cite{newman3,burin3}). Then, in the spirit of the {\it Wheeler-Feynman conjecture}, a set of identical point-like copies of a single particle will be detected by an observer on a single worldline~\footnote{There exists another mechanism to obtain a collection of identical particles on a unique worldline that requires the {\em implicit} definition of the latter. This also leads to a collective conservative algebraic dynamics~\cite{kass8,KhasanMark14}} via his/her light cone (i.e. as the set of solutions of the ``retardation equation''). Remarkably~\cite{KhasanMark15}, for an inertial observer and arbitrary {\it polynomially parameterized} worldline a {\it complete set of Lorentz invariant conservation laws} holds within the emerging collective dynamics, along with some additional properties, such as ``clusterization''. 

However, as mentioned in Sec.~6, the particles-singularities defined by the generating system of equations (the equations of shear-free null geodesic congruence) generally exhibit a much more sophisticated collective {\em string dynamics}, the exploration and elaboration of which will hopefully be undertaken in the future. 
           
\section*{Acknowledgment}
This work was funded by the Ministry of Education and Science of Russia within the 2016-2020 program of support of fundamental research in Peoples' Friendship University.



\begin{thebibliography}{99}


\bibitem{pr}
R. Penrose and W. Rindler, {\it Spinors and Space–Time, vol. 1 \& 2} (Cambridge University Press, Cambridge, 1984)
\bibitem{pr00}R. Penrose, {\it Class. Quant. Grav.}{\bf 14} (1997), A299
\bibitem{LeBrun}LeBrun, {\it Class. Quant. Grav.} {\bf 2} (1985), 555
\bibitem{NewmanSFC}
E.T. Newman, {\it Class. Quant. Grav.} {\bf 26} (2009), 235012, arXiv:0911.4205 [gr-qc]  
\bibitem{Trish}
V.V. Kassandrov and V.N. Trishin, {\it Gen. Rel. Grav.} {\bf 36} (2004), 1603, gr-qc/0401120 
\bibitem{prmc}R. Penrose and M.A.H. MacCallum, {\it Phys. Rep.~C}~{\bf 6} (1972), 241
\bibitem{rob}I. Robinson, {\it J. Math. Phys.} {\bf 2} (1961), 290
\bibitem{NewmanSFMaxwell}
E.T. Newman, {\it Class. Quant. Grav.} {\bf 21} (2004), 3197, gr-qc/0404072
\bibitem{kass9}V.V. Kassandrov, {\it Grav. Cosmol.} {\bf 17} (2011), 47, arXiv:1105.3498 [gr-qc]
\bibitem{dks}G.C. Debney, R.P. Kerr and A. Schild, {\it J. Math. Phys.} {\bf 10} (1969), 1842
\bibitem{demyan}
M. Demianski, {\it Phys. Lett.~A}~{\bf 42} (1972), 157 
\bibitem{burin}
A. Burinskii and G. Magli, {\it Annals Israel Phys. Soc.} {\bf 13} (1997), 296-310, hep-th/9801177 
\bibitem{espos}G. Esposito, {\it Complex General Relativity (Fundamental Theories of Physics, Vol. 69)} (Kluwer, Dordrecht, 1995); gr-qc/9911051 
\bibitem{kr1}V.V. Kassandrov and J.A. Rizcallah, {\it Recent Problems in Field Theory}, ed. A.V. Aminova  (Kasan University Press, Kasan, 1998), gr-qc/9809078 
\bibitem{kr4}V.V. Kassandrov and J.A. Rizcallah, Preprint (2000), gr-qc/0012109
\bibitem{Poten}
G.F. Torres del Castillo, {\it Gen. Rel. Grav.} {\bf 31} (1999), 205  
\bibitem{kass5}V.V. Kassandrov, {\it Quasigroups and Nonassociative
Algebras in Physics}, eds. J. L\~ohmus, P. Kuusk (Institute of Physics
of Estonia' Press, Tallinn, 1990) 
\bibitem{kass1}V.V. Kassandrov, {\it Algebraic Structure of Space-Time and
Algebrodynamics} (Peoples' Friendship University Press, Moscow, 1992) (in Russian)
\bibitem{kass2}V.V. Kassandrov, {\it Grav. Cosmol.} {\bf 3} (1995), 216, gr-qc/0007027
\bibitem{Tod}K.P. Tod, {\it Class. Quant. Grav.} {\bf 13} (1996), 2609
\bibitem{kass3}V.V. Kassandrov, {\it Acta. Applic. Math.} {\bf 50} (1998), 197
\bibitem{kass7}V.V. Kassandrov, {\it Space-Time Structure. Algebra and Geometry}, eds. D.G. Pavlov, Gh. Atanasiu, V. Balan. (Lilia-Print, Moscow, 2007), arXiv:0710.2895 [math-ph] 
\bibitem{YadPhys}
V.V. Kassandrov, {\it Phys. Atom. Nuclei} {\bf 72} (2009), 813, arXiv:0907.5425 [physics. gen-ph]
\bibitem{Gsponer}
A. Gsponer and J-P Hurni, Preprints (2005), math-ph/0510059; math-ph/0511092  
\bibitem{sud}A. Sudbery, {\it Proc. Camb. Phil. Soc.} {\bf 85} (1979), 199
\bibitem{kr2}V.V. Kassandrov and J.A. Rizcallah, Preprint (1998), gr-qc/9809056
\bibitem{Grgin}
E. Grgin, {\it Phys. Lett.~B} {\bf 431} (1998), 15  
\bibitem{kr5}V.V. Kassandrov and J.A. Rizcallah, {\it Gen. Rel. Grav.} {\bf 46} (2014), 1771, arXiv:1311.5423 [gr-qc]
\bibitem{KasJos15} V.V. Kassandrov and J.A. Rizcallah, {\it Grav. Cosmol.} {\bf 21(4)} (2015), 273,  arXiv:1510.01228 [gr-qc]
\bibitem{Mark16}   V.V. Kassandrov and N.V. Markova, {\it Grav. Cosmol.} {\bf 22(4)} (2016), 363
\bibitem{kass4}V.V. Kassandrov, {\it Has the Last Word been Said on Classical Electrodynamics?}, eds. A. Chubykalo, V. Onoochin, A. Espinoza, V. Smirnov-Rueda. (Rinton Press, 2003), physics/0308045 
[physics.class-ph]
\bibitem{kw}R.P. Kerr and W.B. Wilson, {\it Gen. Rel. Grav.} {\bf 10} (1979), 273
\bibitem{diss}J.A. Rizcallah, Geometrization of electromagnetism on the basis of spaces with Weyl-Cartan connections, Ph. D. Thesis, Moscow, 1999 (in Russian)
\bibitem{kt}V.V. Kassandrov and V. N. Trishin, {\it Grav. Cosmol.} {\bf 5} (1999), 272, gr-qc/0007026
\bibitem{ksmh}H. Stephani, D. Kramer, M. MacCallum, C. Hoenselaers and E. Herlt, {\it Exact Solutions of Einstein's Field Equations} (Cambridge University Press, Cambridge, 2003) 
\bibitem{ror}F. Rorlich, {\it Classical Charged Particles} (World Scientific, 2007) 
\bibitem{ranada2}A.F. Ran\~ada, {\it Phys. Lett.~A}~{\bf 310} (2003), 134
\bibitem{zhuravl}V.N. Zhuravlev, {\it Grav. Cosmol.} {\bf 17} (2011), 201
\bibitem{carter}B. Carter, {\it Phys. Rev.} {\bf 174} (1968), 1559
\bibitem{lopes}C.A. Lopez, {\it Phys. Rev.~D }~{\bf 30} (1984), 313
\bibitem{israel}W. Israel, {\it Phys. Rev.~D}~{\bf 2} (1970), 641
\bibitem{burin1}A. Burinskii, {\it Sov. Phys. - JETP} {\bf 39} (1974), 193
\bibitem{burin2}
A. Burinskii, {\it Theor. and Math. Phys.} {\bf 177(2)} (2013), 1492, arXiv:1307.5021 [hep-th]
\bibitem{NEWMANElectron}
 E.T. Newman, {\it Phys. Rev.~D}~{\bf 65} (2002), 104005, gr-qc/0201055 
\bibitem{BURElectron}
A. Burinskii, {\it Grav. Cosmol.} {\bf 14} (2008), 109, hep-th/0507109
\bibitem{newman3}T.M. Adamo and E.T. Newman, {\it Class. Quant. Grav.} {\bf 27} (2010), 075009, arXiv:0911.4205 [gr-qc] 
\bibitem{burin3}A. Burinskii, {\it Phys. Rev.~D}~{\bf 67} (2003), 124024, gr-qc/0212048
\bibitem{kass8}V.V. Kassandrov and I.Sh. Khasanov, {\it Journal of Physics A: Math. Theor.} {\bf 46} (2013), 175206, arXiv:1211.7002 [physics.gen-ph]
\bibitem{KhasanMark14}V.V. Kassandrov, I.Sh. Khasanov and N.V. Markova, {\it Bulletin Peoples' Friendship University (Math. Inform.Sci. Physics)} {\bf 2} (2014), 169; arXiv:1402.6158 [math-ph] 
\bibitem{KhasanMark15}V.V. Kassandrov, I.Sh. Khasanov and N.V. Markova, {\it Journal of Physics A: Math. Theor.} {\bf 48} (2015), 395204; arXiv:1501.01606 [physics.gen-ph]
\end{thebibliography}
\end{document}